\documentclass[seceq]{ptptex}

\usepackage{graphicx}

\def\tr{\mathop{\rm tr}\nolimits}

\newcommand{\ol}{\overline}
\newcommand{\wt}{\widetilde}

\title{
On the Moduli Space of Elliptic Maxwell-Chern-Simons Theories%
}

\author{
Yosuke \textsc{Imamura}\thanks{E-mail: \tt imamura@hep-th.phys.s.u-tokyo.ac.jp}\quad
and\quad
Keisuke \textsc{Kimura}\thanks{E-mail: \tt kimura@hep-th.phys.s.u-tokyo.ac.jp}%
}

\inst{
Department of Physics, The University of Tokyo,
Tokyo 113-0033, Japan
}

\abst{
We analyze the moduli space of
the low-energy limit of 3-dimensional
${\cal N}=3$ Maxwell-Chern-Simons theories
described by circular quiver diagrams, as for
4-dimensional elliptic models.
We define the theories by using D3-NS5-(k,1)5-brane systems
with an arbitrary number of fivebranes.
The supersymmetry is expected to be enhanced to ${\cal N}=4$
in the low-energy limit.
We show that the Higgs branch,
in which all bifundamental scalar fields develop vacuum expectation values,
is an abelian orbifold of ${\mathbb C}^4$.
We confirm that the same geometry is obtained
as an M-theory dual of the brane system.
We also consider theories realized by
introducing more than two kinds of fivebranes,
and obtain nontoric fourfolds as moduli spaces.
}

\begin{document}

\maketitle

\section{Introduction}
Recently, there has been great interest in
3-dimensional superconformal field theories
as theories for describing multiple M2-branes in various backgrounds.
This was triggered by the proposal of a new class of
3-dimensional theories by
Bagger and Lambert\cite{Bagger:2006sk,Bagger:2007jr,Bagger:2007vi},
and Gusstavson\cite{Gustavsson:2007vu,Gustavsson:2008dy}.
The model (BLG model) possesses
${\cal N}_{(d=3)}=8$ superconformal symmetry
and is based on Lie $3$-algebra.
The action of the BLG model includes
the structure constant $f^{abc}{}_d$ of a Lie $3$-algebra,
which determines the form
of the interactions, and a metric $h^{ab}$,
which appears in the coefficients of the kinetic terms.
These tensors must satisfy certain conditions required by the
supersymmetry invariance of the action.
If these tensors satisfy the conditions,
we can write down the action of a BLG model.
The constraint imposed on the structure constant
is called a fundamental identity.
It was soon realized that the identity is
very restrictive\cite{Ho:2008bn},
and it was proved that
if we assume that the metric is positive definite
and the algebra is finite dimensional,
there is only one nontrivial Lie $3$-algebra\cite{Papadopoulos:2008sk,Gauntlett:2008uf}, which is called an $A_4$ algebra.
The BLG model based on the $A_4$ algebra
is a $SU(2)\times SU(2)$ Chern-Simons theory
with levels $k$ and $-k$ for each $SU(2)$ factor.
Analysis of this model
showed that it describes a pair of M2-branes
in certain orbifold backgrounds\cite{VanRaamsdonk:2008ft,Lambert:2008et,Distler:2008mk}.
As a theory for an arbitrary number of M2-branes,
a model based on an algebra with
a Lorenzian metric was proposed in
Refs.~\citen{Gomis:2008uv,Benvenuti:2008bt,Ho:2008ei}.
Because of the indefinite metric,
the model includes unwanted ghost modes.
Although the ghost modes can be removed
by treating them as
background fields satisfying classical equations
of motion\cite{Ho:2008ei,Honma:2008un},
or by gauging certain symmetries and fixing them\cite{Bandres:2008kj,Gomis:2008be},
this procedure breaks the conformal invariance,
and the theory becomes D2-brane theory\cite{Ho:2008ei,Bandres:2008kj,Ezhuthachan:2008ch}
by the mechanism proposed in Ref.~\citen{Mukhi:2008ux}
unless the parameter corresponding to the Yang-Mills coupling
is sent to infinity or integrated over all values as a dynamical parameter\cite{Gomis:2008be}.

There has also been some progress in
3-dimensional Chern-Simons theories
with supersymmetries of less than $8$,
which are closely related to M2-branes.
Gaiotto and Witten\cite{Gaiotto:2008sd}
proposed ${\cal N}_{(\rm d=3)}=4$
superconformal Chern-Simons theories,
and Hosomichi et al.\cite{Hosomichi:2008jd}
extended the theories
by introducing twisted hypermultiplets.
They derived the relation between their models
and the BLG model,
and showed that the $A_4$ BLG model is included as
a special case of their ${\cal N}_{(d=3)}=4$ Chern-Simons theories.
They also studied the M-crystal model
\cite{Lee:2006hw,Lee:2007kv,Kim:2007ic},
which is described by a circular quiver diagram with $2n$ vertices.
The vertices represent Chern-Simons
fields at level $\pm k$ with alternate
signatures, and by analyzing the moduli space of
the model they showed that it can be regarded as
a theory describing
M2-branes in the orbifold $({\mathbb C}^2/{\mathbb Z}_n)^2$.\footnote{%
The possibility that the existence of magnetic monopoles
causes a discrete indentification in the orbifold is also mentioned.}
They also presented a realization of this model
by using D3-, D5-, and NS5-branes, which give the
model at level $\pm1$, and reproduce the
orbifold as the M-theory dual of the brane system.

Aharony et al. also proposed
a similar model\cite{Aharony:2008ug}
based on $U(N)\times U(N)$ Chern-Simons theory
with levels $k$ and $-k$ for each $U(N)$ factor.
They showed that
the action possesses ${\cal N}_{(d=3)}=6$ superconformal symmetry,
and describes $N$ M2-branes in the
orbifold ${\mathbb C}^4/{\mathbb Z}_k$.
Although ${\cal N}_{(d=3)}=8$ supersymmetry,
which is expected when $k=1$ or $2$ is not manifest,
the action does not have dimensionful parameters and
the scale invariance is manifest.
In Ref.~\citen{Aharony:2008ug} it is also shown that
the theory can be realized as a theory on a
brane system in type IIB string theory.
The brane system consists of $N$ D3-, one NS5-, and
one $(k,1)$5-branes.
They showed that by T-duality and M-theory lift,
M2-branes in the orbifold ${\mathbb C}^4/{\mathbb Z}_k$ are obtained.

The purpose of this paper is to extend
the models proposed in Refs.~\citen{Hosomichi:2008jd} and \citen{Aharony:2008ug}
by generalizing the brane configurations
in these references.
In \S\ref{brane.sec} we consider a
brane system with $n_A$ NS5-branes and $n_B$ $(k,1)$5-branes,
and analyze the moduli space of the theory
realized by the brane system.
The theory is a $U(N)^{n_A+n_B}$
quiver gauge theory
with nonvanishing Chern-Simons terms for some of
the $U(N)$ factors.
Some of the $U(N)$ fields are Yang-Mills fields without
Chern-Simons coupling.
The supersymmetry of this theory is ${\cal N}_{(d=3)}=3$,
which is expected to be enhanced to ${\cal N}_{(d=3)}=4$
in the strong gauge-coupling limit.
The reason for this is as follows.
This theory can be obtained
from the $U(N)\times U(N)$ theory proposed in Ref.~\citen{Gaiotto:2008sd}
by combining two extensions.
One is the inclusion of twisted hypermultiplets, as mentioned above,
and the other is the inclusion of gauge groups
with vanishing Chern-Simons couplings.
The latter extension is discussed in Ref.~\citen{Gaiotto:2008sd} to describe
general nonlinear sigma models of hypermultiplets.
Both extensions are known to give ${\cal N}_{(d=3)}=4$ supersymmetric
Chern-Simons theory,
and it is plausible that the theory we discuss in this paper
possesses ${\cal N}_{(d=3)}=4$ supersymmetry.

In \S\ref{moduli.sec} we determine the moduli
space of the theory.
We focus only on the Higgs branch,
which describes a mobile M2-brane.
Under a certain assumption for flux quantization,
we obtain a 4-dimensional orbifold ${\mathbb C}^4/\Gamma$,
where $\Gamma$ is a discrete subgroup
depending on $k$, $n_A$, and $n_B$.
We reproduce the same orbifold in \S\ref{dual.sec}
as an M-theory dual of the brane configuration.
In \S\ref{more.sec} we consider models with more than two kinds of fivebranes.
The moduli space is also a 4-dimensional manifold,
but it is nontoric.
The last section is devoted to discussion.

\section{Brane configuration and action}\label{brane.sec}
The model proposed in Ref.~\citen{Aharony:2008ug}
is a Chern-Simons theory
with a $U(N)\times U(N)$ gauge group.
It can be realized as a theory based on a brane system
consisting of $N$ D3-branes, one NS5-brane, and one $(k,1)$5-brane.
All these branes share the directions of 012, which are the coordinates of the
3-dimensional field theory.
The $N$ D3-branes are wrapped on the compact direction 9.
The NS5-brane and the $(k,1)$5-brane are spread along
the 012345 and $012[36]_{\theta_1}[47]_{\theta_2}[58]_{\theta_3}$ directions,
respectively, where $[ij]_\theta$ is the direction in the $i$-$j$ plane
specified by the angle $\theta$.
The angles $\theta_{1,2,3}$ are determined by the BPS conditions.
\begin{figure}[t]
\centerline{\includegraphics{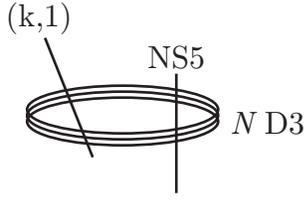}}
\caption{Brane configuration for the $U(N)\times U(N)$ Chern-Simons model.}
\label{brane.eps}
\end{figure}
We refer to NS5- and $(k,1)$5-branes as A- and B-branes, respectively.
The D3-brane worldvolume is divided into two parts by the intersecting fivebranes (Fig.~\ref{brane.eps}),
and a $U(N)$ vector multiplet exists on each segment.
Bifundamental chiral multiplets also arise
at the intersections.
This brane system is similar to the D4-NS5 system
realizing the Klebanov-Witten theory\cite{Klebanov:1998hh},
which is a 4-dimensional ${\cal N}_{(d=4)}=1$ superconformal
field theory.
In the D4-NS5 system, we have $N$ D4-branes wrapped on ${\mathbb S}^1$,
instead of D3-branes,
and the A- and B-branes in this case are
NS5-branes along different directions.

We generalize the D3-fivebrane system by introducing an arbitrary number of
fivebranes.
In the case of 4-dimensional ${\cal N}_{(d=4)}=1$ gauge theories,
such a generalization is known as an elliptic model,
and has been studied in detail\cite{Uranga:1998vf,vonUnge:1999hc}.
It is known that the moduli spaces of the
theories are generalized conifolds.
We here carry out a similar analysis in the 3-dimensional case.
Let $n_A$ and $n_B$ be the numbers of A- and B-branes, respectively.
We denote the total number of fivebranes by $n=n_A+n_B$.
Let us label the fivebranes by $I=1,\ldots,n$
according to their order along ${\mathbb S}^1$.
We identify $I=n+1$ with $I=1$.
On the interval of D3-branes between two fivebranes
$I$ and $I+1$, we have a $U(N)$ vector multiplet $V_I$
and an adjoint chiral multiplet $\Phi_I$.
(We use the terminology of ${\cal N}_{(d=4)}=(1/2){\cal N}_{(d=3)}=1$ supersymmetry.)
The kinetic terms of these multiplets are
\begin{eqnarray}
S_V&=&\int d^3x\sum_I\frac{1}{g_I^2}\tr \left[
-\frac{1}{4}(F^I_{\mu\nu})^2-\frac{1}{2}(D_\mu\sigma_I)^2+\frac{1}{2}D_I^2
+\mbox{fermions}\right],
\label{sv}\\
S_\Phi&=&\int d^3xd^4\theta\sum_I \frac{1}{g_I^2}\tr (\Phi_I^*e^{V_I}\Phi_Ie^{-V_I}).\label{sphi}
\end{eqnarray}
$\sigma_I$ is the real scalar field in the vector multiplet $V_I$.
The adjoint chiral multiplets $\Phi_I$ describe the
motion of the D-branes along the fivebranes.
When two fivebranes $I$ and $I+1$ are not parallel,
the chiral multiplet $\Phi_I$ becomes massive,
and the mass term is described by the superpotential
\begin{equation}
W=\frac{\mu}{2}\sum_I(q_{I+1}-q_I)\Phi_I^2,
\label{massterm}
\end{equation}
where $q_I=0$ for A-branes and $q_I=1$ for B-branes.
The overall factor $\mu$ is related to the relative angle
between A- and B-branes.

\begin{figure}[t]
\centerline{\includegraphics{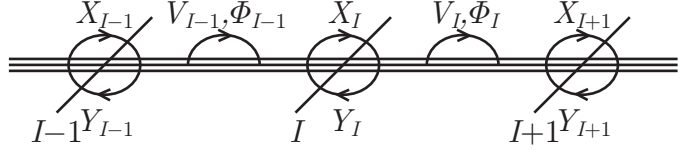}}
\caption{Brane system and fields.}
\label{xyfv.eps}
\end{figure}

We also have bifundamental chiral multiplets $X_I$ and $Y_I$, which arise from
open strings stretched between two intervals of D-branes
divided by the $I$th fivebrane.
(See Fig.~\ref{xyfv.eps}.)
These fields belong to the following representations of
$U(N)_I\times U(N)_{I-1}$, where $U(N)_I$ is the gauge group
associated with the vector multiplet $V_I$:
\begin{equation}
X_I:(N,\ol N),\quad
Y_I:(\ol N,N).
\end{equation}
The kinetic terms of these bifundamental fields are
\begin{eqnarray}
S_{XY}
&=&\int d^3x d^4\theta\sum_{I=1}^n\tr\left[
X_I^*e^V_IX_Ie^{-V_{I-1}}
+Y_Ie^{-V_I}Y^*_Ie^{V_{I-1}}
\right]
\nonumber\\
&=&\int d^3x
\sum_{I=1}^n\tr\left[
 -D_I(|X_I|^2-|Y_I|^2-|X_{I+1}|^2+|Y_{I+1}|^2)
\right.
\nonumber\\&&
\left.
 -(|X_I|^2+|Y_I|^2)(\sigma_I-\sigma_{I-1})^2
  +|F^X_I|^2
  +|F^Y_I|^2
 \right]
+\cdots.
\end{eqnarray}
In the component expression we show only the bosonic terms without derivatives.
These bifundamental fields couple
to the adjoint chiral multiplets
through the superpotential
\begin{equation}
W=\sum_{I=1}^n\tr\Phi_I(X_IY_I-Y_{I+1}X_{I+1}).
\label{n2term}
\end{equation}

The difference between the RR-charges of the A- and B-branes
generates Chern-Simons terms\cite{Kitao:1998mf,Bergman:1999na}.
The bosonic part of
the ${\cal N}_{d=3}=2$ completion of the Chern-Simons terms
is
\begin{equation}
S_{\rm CS}
=\sum_{I=1}^n\frac{k_I}{2\pi}\int d^3x\tr\left[
\epsilon^{\mu\nu\rho}
\left(\frac{1}{2}A^I_\mu\partial_\nu A^I_\rho
+\frac{1}{3}A^I_\mu A^I_\nu A^I_\rho\right)
+\sigma_I D_I
\right],
\label{csterm}
\end{equation}
where the Chern-Simons coupling $k_I$ is given by
\begin{equation}
k_I=k(q_{I+1}-q_I).
\label{cscouplings}
\end{equation}
We assume that $k$ is a positive integer.
The Chern-Simons terms in (\ref{csterm})
cause some of the vector multiplets to be
massive.
The masses $\sim k_Ig_I^2$
are proportional to the masses of adjoint chiral multiplets $\Phi_I$.
We can promote the supersymmetry of this theory to ${\cal N}_{(d=3)}=3$ by
matching the masses of $V_I$ and $\Phi_I$
by setting $\mu=k$.

In $3$-dimensional field theories the coupling constants $g_I$ have mass dimension $1/2$, and
taking the low-energy limit
is equivalent to taking the strong-coupling limit
$g_I\rightarrow\infty$.
This makes the masses of $V_I$ and $\Phi_I$ infinity
unless $k_I=0$,
and we can integrate out the
massive adjoint chiral multiplets.
After this,
the superpotential becomes\footnote{We shift the field $\Phi_I$ by $(q_I-1/2)(X_IY_I+Y_{I+1}X_{I+1})$
and set $\mu=1$ to simplify the equations.}
\begin{equation}
W=\sum_{q_I=q_{I+1}}\tr\Phi_I(X_IY_I-Y_{I+1}X_{I+1})
+\sum_{q_I\neq q_{I+1}}(q_{I+1}-q_I)\tr(X_IY_IY_{I+1}X_{I+1}).
\label{ellw}
\end{equation}

\section{Moduli space}\label{moduli.sec}
In this section we investigate the moduli space of the
3-dimensional field theory defined in the
previous section.
As we mentioned at the end of the previous section
we need to take the strong-coupling limit $g_I\rightarrow\infty$
to obtain the conformal theory describing the low-energy limit
of M2-branes.
Although the dynamics in such a strong coupling region is
highly nontrivial,
we assume that the vacuum structure is not affected by
quantum corrections,
and we consider only the classical equations of motion
derived from the action given in the previous section.
In the strong-coupling limit, the kinetic terms
(\ref{sv}) and (\ref{sphi}) vanish,
and the fields $\phi_I$, the scalar components of
$\Phi_I$, and $\sigma_I$ become auxiliary fields.
The bifundamental chiral multiplets $X_I$ and $Y_I$ are still dynamical,
and the moduli space is parameterized by
the scalar components of these multiplets.

We are interested in the moduli space for a single M2-brane, and we set $N=1$.
Furthermore, we here focus only on the Higgs branch,
which describes a mobile M2-brane,
and assume
\begin{equation}
X_I,
Y_I\neq0.
\label{coulomb}
\end{equation}

\subsection{F-term conditions}
Let us first consider the F-term conditions
derived from the superpotential (\ref{ellw}).
Because the superpotential
is the same as the 4-dimensional elliptic model
realized by the D4-NS5 brane system,
the F-term conditions are also the same.
Under the assumption (\ref{coulomb}),
the F-term conditions give the following solution:
\begin{equation}
\Phi_{I\in A}=M_{I\in B}=u,\quad
\Phi_{I\in B}=M_{I\in A}=v,
\label{fcond}
\end{equation}
where we define the mesonic operators as $M_I=X_IY_I$.
$I\in A$ ($I\in B$) means that index $I$ is restricted
to the values with $q_I=0$ ($q_I=1$).

\begin{figure}[t]
\centerline{\includegraphics{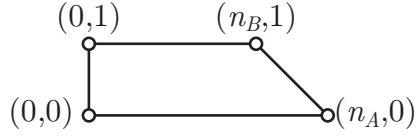}}
\caption{Toric diagram of a generalized conifold.}
\label{gc.eps}
\end{figure}

Although not directly related to our model,
it may be instructive to demonstrate how we can obtain a Calabi-Yau $3$-fold as the moduli space
of a 4-dimensional elliptic model
in the case of the D4-NS5 system.
In this case two complex numbers $u$ and $v$ can be interpreted as the
coordinates of the D4-brane along B- and A-branes, respectively.
The 4-dimensional theory possesses $U(1)^{n-1}$ gauge symmetry.
In addition to the mesonic operators $M_I$,
we can construct the gauge-invariant
baryonic operators
\begin{equation}
x=\prod_{I=1}^nX_I,\quad
y=\prod_{I=1}^nY_I.
\label{baryon}
\end{equation}
By definition, these gauge-invariant operators are related by
\begin{equation}
xy=u^{n_A} v^{n_B}.
\end{equation}
This algebraic equation defines
a Calabi-Yau 3-fold, which is often called a generalized conifold.
The toric diagram of this generalized conifold is shown in Fig.~\ref{gc.eps}.

\subsection{D-term conditions}\label{dterm.sec}
In the strong-coupling limit $g_I\rightarrow\infty$,
the vector multiplet $V_I$ includes two auxiliary fields $\sigma_I$ and $D_I$.
The terms in the action including these auxiliary fields are
\begin{eqnarray}
S&=&\sum_{I=1}^n\left[ k_I\sigma_ID_I
 -D_I(|X_I|^2-|Y_I|^2-|X_{I+1}|^2+|Y_{I+1}|^2)
\right.
\nonumber\\&&
\left.
 -(|X_I|^2+|Y_I|^2)(\sigma_I-\sigma_{I-1})^2
\right].
\label{daction}
\end{eqnarray}
In this action, $D_I$ are Lagrange multipliers,
and give the constraint
\begin{equation}
k_I\sigma_I=|X_I|^2-|Y_I|^2-|X_{I+1}|^2+|Y_{I+1}|^2.
\label{Dconst}
\end{equation}
If we substitute this into the 
action (\ref{daction}),
the first line vanishes and the potential becomes
\begin{equation}
V=\sum_{I=1}^n(|X_I|^2+|Y_I|^2)(\sigma_I-\sigma_{I-1})^2.
\end{equation}
Because of the assumption (\ref{coulomb}),
vacua are given by $\sigma_I=\sigma_{I-1}$.
Namely, all $\sigma_I$ are the same.
Let $\sigma$ be the common value of $\sigma_I$.
Then the constraint (\ref{Dconst}) becomes
\begin{equation}
q_I\sigma-(|X_I|^2-|Y_I|^2)
=q_{I+1}\sigma-(|X_{I+1}|^2-|Y_{I+1}|^2).
\end{equation}
This means that the left- and right-hand sides of this
equation do not depend on the index $I$.
Thus, we can write
\begin{equation}
|X_I|^2-|Y_I|^2=q_I\sigma+c
\label{dterm}
\end{equation}
with a constant $c$.

Although (\ref{dterm})
is not the equation of motion of $D_I$,
we can formally interpret it
as an ordinary D-term condition associated with a certain symmetry.
To rewrite (\ref{dterm}) in the form
of an ordinary D-term condition,
let us define $U(1)$ transformation groups $G_I$ that act only
on $X_I$ and $Y_I$ as
\begin{equation}
G_I:
X_I\rightarrow e^{i\lambda_I}X_I,\quad
Y_I\rightarrow e^{-i\lambda_I}Y_I,
\end{equation}
where $\lambda_I$ is a parameter of $G_I$.
The groups $G_I$ are different from $U(1)_I$ defined in the previous section.
The parameters $\alpha_I$ of $U(1)_I$ and $\lambda_I$ of $G_I$ are
related by
\begin{equation}
\lambda_I=\alpha_I-\alpha_{I-1}.
\label{lambdaalpha}
\end{equation}
Although each $G_I$ is not a symmetry of the theory,
it is convenient to
describe symmetry groups as subgroups of
$\prod_IG_I$.
For example,
the gauge symmetry $G=U(1)^{n-1}$,
which does not include the diagonal $U(1)$ decoupling
from the theory,
is the subgroup of $\prod_IG_I$ that
does not rotate the baryonic operators (\ref{baryon}).

Let us rewrite (\ref{dterm}) in the form of a D-term condition.
Equation (\ref{dterm}) is equivalent to
the condition
\begin{equation}
\sum_{I=1}^l\lambda_I(|X_I|^2-|Y_I|^2)=0,
\label{dtermc}
\end{equation}
for arbitrary $\lambda_I$ satisfying the constraints
\begin{equation}
\sum_{I=1}^n\lambda_I=\sum_{I=1}^nq_I\lambda_I=0.
\label{subg}
\end{equation}
If we regard $\lambda_I$ as
the parameters of $G_I$ transformations,
the constraints (\ref{subg}) imposed on
$\lambda_I$ define a subgroup $H=U(1)^{n-2}$ of $\prod_I G_I$.
Equation (\ref{dtermc}) can be regarded as the D-term condition
for $H$.

We emphasize that we do not claim at this point
that the gauge symmetry
of the theory is $H$ or that relation (\ref{dterm}) is
obtained as the equations of motion of auxiliary fields
in the vector multiplets associated with $H$.
We only claim that the vacuum condition
(\ref{dterm}) is similar to the D-term condition
of a gauge theory with the gauge symmetry $H$.
In the next subsection,
however, we will show that $H$ indeed emerges as the
unbroken continuous gauge symmetry.

It is convenient to define
the subgroup $H$ in another way.
Let us define the baryonic operators
\begin{equation}
x_A=\prod_{I\in A}X_I,\quad
y_A=\prod_{I\in A}Y_I,\quad
x_B=\prod_{I\in B}X_I,\quad
y_B=\prod_{I\in B}Y_I.
\label{baryonicops}
\end{equation}
The group $H$ can be defined
as the subgroup of $\prod_IG_I$ that does not rotate these baryonic operators.

\subsection{Gauge symmetry}
To obtain the moduli space of a gauge theory,
we need to remove unphysical degrees of freedom
corresponding to gauge symmetries.
In the case of Chern-Simons theories,
we should carefully take account of symmetry breaking due to
the existence of magnetic monopoles.
Let us rewrite the abelian Chern-Simons terms
in the form
\begin{equation}
S_{\rm CS}
=-\frac{k}{2\pi}\sum_{I=1}^nq_I(A^I-A^{I-1})\wedge \wt F
 +(\mbox{quadratic terms of $A^I-A^{I-1}$}),
\label{csdiag}
\end{equation}
where $\wt F$ is the field strength of the diagonal
$U(1)$ gauge field $\wt A=(1/n)(A^1+A^2+\cdots+A^n)$.
Equation (\ref{csdiag}) is obtained by substituting
\begin{equation}
A^I=\wt A+(\mbox{linear combination of $A^I-A^{I-1}$})
\end{equation}
into the Chern-Simons term in (\ref{csterm}).
The quadratic term of $\wt A$ vanishes because $\sum_Ik_I=0$.
Because the diagonal gauge field $\wt A$ appears only in the
first term of (\ref{csdiag}),
we can dualize it by adding the term
\begin{equation}
\frac{1}{2\pi}\int d\tau\wedge\wt F,
\end{equation}
and treating $\wt F$ as an unconstrained field.
The equation of motion of $\wt F$ gives
\begin{equation}
\sum_{I=1}^n k_IA_I=d\tau.
\end{equation}
Upon the gauge transformation $\delta A_I=d\alpha_I$,
the scalar field $\tau$ is transformed as
\begin{equation}
\delta \tau=\sum_{I=1}^n k_I\alpha_I.
\end{equation}
Let us assume that the period of $\tau$ is $2\pi$.
This implies that the flux $\oint \wt F$ is quantized by
\begin{equation}
\int\wt F\in 2\pi{\mathbb Z}.
\label{feqf2}
\end{equation}
Although we could not show this flux quantization
on the field-theory side,
we will later show that the moduli space obtained
by assuming (\ref{feqf2}) coincides with
that obtained from the brane configuration
by the T-duality and M-theory lift.
If we adopt this assumption,
the gauge fixing $\tau=0$ partially breaks the gauge symmetry
and imposes the following constraint
on the parameters $\lambda_I$ and $\alpha_I$:
\begin{equation}
\sum_{I=1}^n k_I\alpha_I=k\sum_{I=1}^nq_I\lambda_I\in 2\pi{\mathbb Z}.
\label{ubk}
\end{equation}
(In the first equality we used (\ref{cscouplings}) and (\ref{lambdaalpha}).)

Let us first focus on the continuous subgroup.
It is generated by parameters satisfying
\begin{equation}
\sum_{I=1}^n\lambda_I=\sum_{I=1}^nq_I\lambda_I=0.
\label{sumqlambda}
\end{equation}
The group defined by (\ref{sumqlambda}) is simply
group $H$ defined in \S\ref{dterm.sec}.
Because of the emergence of the same group $H$
both in the equations of motion of auxiliary fields
and in the unbroken gauge symmetry,
we can obtain the moduli space
as the coset ${\cal M}/H_{\mathbb C}$ or its orbifold,
where ${\cal M}$ is the complex manifold defined by
the F-term conditions and $H_{\mathbb C}$ is
the complexification of the group $H$.
This guarantees that the moduli space is a complex manifold.

In addition to $H$,
the group defined by (\ref{ubk})
includes the discrete symmetry, which rotates the baryonic
operators in (\ref{baryonicops}) as
\begin{equation}
x_A\rightarrow e^{\frac{2\pi i}{k}}x_A,\quad
y_A\rightarrow e^{-\frac{2\pi i}{k}}y_A,\quad
x_B\rightarrow e^{-\frac{2\pi i}{k}}x_B,\quad
y_B\rightarrow e^{\frac{2\pi i}{k}}y_B.
\label{discrete}
\end{equation}

\subsection{Moduli space}\label{moduli.ssec}
Let us determine the moduli space.
We first consider the $k=1$ case.
In this case, the discrete gauge symmetry 
(\ref{discrete}) becomes trivial, and
we have the gauge-invariant operators
\begin{equation}
u,\quad
v,\quad
x_A,\quad
y_A,\quad
x_B,\quad
y_B.
\label{giops}
\end{equation}
By definition, these operators satisfy
the following equations:
\begin{equation}
x_Ay_A=u^{n_A},\quad
x_By_B=v^{n_B}.
\label{defeq}
\end{equation}
These equations define the orbifold
${\mathbb C}^2/{\mathbb Z}_{n_A}\times{\mathbb C}^2/{\mathbb Z}_{n_B}$.
Actually, the relation (\ref{defeq}) can be
solved as
\begin{equation}
x_A=z_1^{n_A},\quad
y_A=z_2^{n_A},\quad
u=z_1z_2,\quad
x_B=z_3^{n_B},\quad
y_B=z_4^{n_B},\quad
v=z_3z_4.
\label{zdef}
\end{equation}
We can identify $z_i$ as the coordinates of ${\mathbb C}^4$, the covering space
of the orbifold.
None of the variables in (\ref{giops}) are changed
by the transformations
\begin{equation}
(z_1,z_2,z_3,z_4)\rightarrow
(e^{2\pi i/n_A}z_1,
e^{-2\pi i/n_A}z_2,
z_3,z_4)
\label{zna}
\end{equation}
and
\begin{equation}
(z_1,z_2,z_3,z_4)\rightarrow
(z_1,z_2,
e^{2\pi i/n_B}z_3,
e^{-2\pi i/n_B}z_4).
\label{znb}
\end{equation}
Points in ${\mathbb C}^4$ mapped by these transformations
should be identified with each other, and this
identification defines the above orbifold.

If $n_A=n_B$, the moduli space agrees with the result in Ref.~\citen{Hosomichi:2008jd},
in which alternate A- and B-branes are considered.
It is interesting that the moduli space does not
depend on the order of the two kinds of fivebranes.

Next, let us consider the case when $k>1$.
In this case, we should take account of the
discrete gauge transformation (\ref{discrete}).
The transformation of $z_i$ reproducing (\ref{discrete}) is
\begin{equation}
(z_1,z_2,z_3,z_4)\rightarrow
(e^{2\pi i/kn_A}z_1,e^{-2\pi i/kn_A}z_2,e^{-2\pi i/kn_B}z_3,e^{2\pi i/kn_B}z_4).
\label{znc}
\end{equation}
The three transformations
(\ref{zna}), (\ref{znb}), and (\ref{znc})
generate a discrete subgroup of $U(1)^2$ with
$kn_An_B$ elements.
Let $\Gamma$ be this discrete group.
The moduli space for general $k$ is
the abelian orbifold ${\mathbb C}^4/\Gamma$.

\section{M-theory dual}\label{dual.sec}
In the previous section,
we obtained the 4-dimensional orbifold ${\mathbb C}^4/\Gamma$
as the Higgs branch of the moduli space.
The purpose of this section is to reproduce the same
orbifold by
the T-duality transformation
and the M-theory lift from the D3-fivebrane system in type IIB string theory.

For simplicity, we first consider a system in which the
$(k,1)$5-branes are replaced by D5-branes.
After determining the mapping from type IIB string theory to
M-theory for NS5- and D5-branes,
the dual object for the bound state of these two kinds of branes
is easily obtained by superposing the objects for NS5- and D5-branes.
Although the tilted angle of the branes should be
appropriately chosen according to the charges of branes
to preserve supersymmetry,
we do not do this because the toric data
do not change upon continuous deformations of the manifold and
because we can determine the toric data of the dual geometry
by using only the topological information.
We start from the brane configuration for type IIB string theory in Table
\ref{d5ns5.tbl}.
\begin{table}[b]
\caption{Brane configuration in type IIB string theory}
\label{d5ns5.tbl}
\begin{center}
\begin{tabular}{c|ccc|ccc|ccc|c}
& 0 & 1 & 2 & 3 & 4 & 5 & 6 & 7 & 8 & 9 \\
\hline
D3 & $\circ$ & $\circ$ & $\circ$ & & & &&&& $\circ$ \\
D5 & $\circ$ & $\circ$ & $\circ$ & &&& $\circ$ & $\circ$ & $\circ$ \\
NS5 & $\circ$ & $\circ$ & $\circ$ & $\circ$ & $\circ$ & $\circ$
&&&\\
\end{tabular}
\end{center}
\end{table}
Direction $9$ is compactified on ${\mathbb S}^1$.
We replaced the $(k,1)$5-brane with the D5-brane and use a coordinate system
in which the D5-brane is spread along 012678.
In general, the $(k,1)$5-branes
are not perpendicular to the NS5-brane,
thus we use slanted coordinates.

We first rearrange the coordinates in
$4578$ space by using the Hopf fibration.
We define $r_a$ ($a=1,2,3$) by
\begin{equation}
r_a=u^\dagger \sigma_au,\quad
u=\left(\begin{array}{c}x^4+ix^5\\x^7+ix^8\end{array}\right),
\end{equation}
and we let $\psi$ be the coordinate of the ${\mathbb S}^1$ fiber.
Then the NS5 and D5 worldvolumes are on the positive and negative parts
of the $r_3$ axis, respectively, in the $r_a$ space.
See Table \ref{d5ns52.tbl}.
The $\psi$ cycle shrinks at the center of the $r_a$ space, 
which is shown in the table as ``KKM''.
``s'' in the table represents the shrinking cycle.
\begin{table}[b]
\caption{The same configuration as Table \ref{d5ns5.tbl} with
a different coordinate system.
``s'' represents the shrinking cycle and $+$ and $-$
mean that the branes are spread along the positive or negative part
of the axis, respectively.}
\label{d5ns52.tbl}
\begin{center}
\begin{tabular}{c|ccc|ccc|cc|cc}
& 0 & 1 & 2 & 3 & 6 & $r_3$ & $r_1$ & $r_2$ & $\psi$ & 9 \\
\hline
D3 & $\circ$ & $\circ$ & $\circ$ &&& &&&&$\circ$\\
D5 & $\circ$ & $\circ$ & $\circ$ && $\circ$ & $-$ &&& $\circ$\\
NS5 & $\circ$ & $\circ$ & $\circ$ & $\circ$ && $+$ &&&$\circ$\\
KKM & $\circ$ & $\circ$ & $\circ$ & $\circ$ & $\circ$ & &&&  s & $\circ$\\
\end{tabular}
\end{center}
\end{table}

Let us perform the T-duality transformation along direction $9$,
and lift the configuration into M-theory.
\begin{table}[t]
\caption{M-theory dual of the brane configuration.}
\label{d5ns53.tbl}
\begin{center}
\begin{tabular}{c|ccc|ccc|cc|ccc}
& 0 & 1 & 2 & 3 & 6 & $r_3$ & $r_1$ & $r_2$ & $\psi$ & 9 & M \\
\hline
(D3$\rightarrow$)M2 & $\circ$ & $\circ$ & $\circ$ &&&&&&&\\
(D5$\rightarrow$)KKM & $\circ$ & $\circ$ & $\circ$ && $\circ$ & $-$ &&& $\circ$ & s & $\circ$ \\
(NS5$\rightarrow$)KKM & $\circ$ & $\circ$ & $\circ$ & $\circ$ && $+$ &&&$\circ$ & $\circ$ & s \\
KKM & $\circ$ & $\circ$ & $\circ$ & $\circ$ & $\circ$ & &&&  s & $\circ$ & $\circ$ \\
\end{tabular}
\end{center}
\end{table}
The D3-branes are mapped to M2-branes as shown in Table~\ref{d5ns53.tbl}.
A single NS5-brane and a single D5-brane
become KKM-branes associated
with the $(1,0,0)$ and $(0,1,0)$ cycles, respectively,
where the first, second, and last components
correspond to the $M$, $9$, and $\psi$ coordinates,
respectively.
If we start with a $(k,1)$5-brane, which is the bound state of
$k$ D5-branes and one NS5-brane,
we obtain a single KKM-brane with
$(0,1,k)$ cycle shrinking.

\begin{figure}[b]
\centerline{\includegraphics{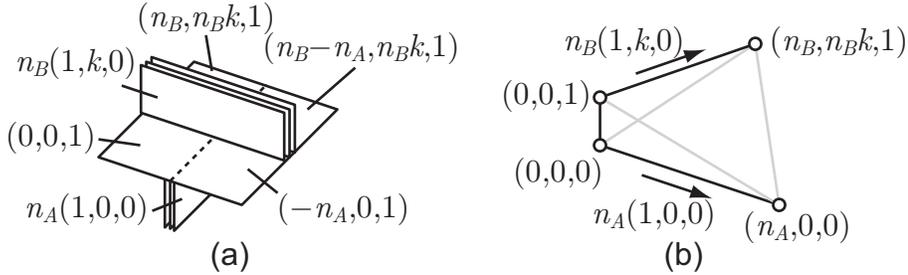}}
\caption{(a) M-theory dual of the D3-fivebrane system in $36r_3$ space.
This can be regarded as a webdiagram of the toric geometry.
The corresponding toric diagram is shown in (b).}
\label{c4.eps}
\end{figure}

In addition to these, we have one more KKM-brane, which originates
from the special choice of the coordinates.
The existence of the other KKM-branes
make the shrinking cycle of the last KKM-brane ambiguous,
and only the last
component of the shrinking cycle has a definite value of $1$.
The intersection with other branes changes the
shrinking cycle, and the cycle should be determined
according to the ``charge conservation''
of the KKM branes.
See Fig.~\ref{c4.eps}(a).

This system of KKM-branes in $36r_3$ space is simply a webdiagram
describing a 4-dimensional toric manifold.
We can easily obtain the toric diagram as a dual graph of the
webdiagram. (Fig.~\ref{c4.eps}(b))
The toric variety described by this diagram is in fact the
orbifold we obtained in \S\ref{moduli.ssec},
as we show in the rest of this section.

The structure of a toric variety is mostly determined by
the toric data, which are a set of
generators of shrinking cycles.
The generators are usually
represented as vectors $\vec v_i$ in the
lattice associated with the toric fiber.
The toric data of ${\mathbb C}^4$
are given by $\vec v_i=\vec e_i$ ($i=1,2,3,4$), where
$\vec e_i$ are the unit vectors in the 4-dimensional lattice.
\begin{equation}
\vec e_1=(1,0,0,0),\quad
\vec e_2=(0,1,0,0),\quad
\vec e_3=(0,0,1,0),\quad
\vec e_4=(0,0,0,1).
\end{equation}
The orbifolding of a toric variety is
realized by refining the lattice
by adding new generators.
In the case of the orbifold defined by (\ref{zna})--(\ref{znc}),
we add three generators
\begin{eqnarray}
\vec e_5&=&\left(\frac{1}{n_A},-\frac{1}{n_A},0,0\right),\nonumber\\
\vec e_6&=&\left(0,0,\frac{1}{n_B},-\frac{1}{n_B}\right),\nonumber\\
\vec e_7&=&\left(\frac{1}{n_Ak},-\frac{1}{n_Ak},-\frac{1}{n_Bk},\frac{1}{n_Bk}\right).
\end{eqnarray}
Of course, the seven vectors $\vec e_1,\ldots,\vec e_7$ are
not linearly independent.
Let us choose the following linearly independent basis:
\begin{eqnarray}
\vec f_1&=&-\vec e_5=\left(-\frac{1}{n_A},\frac{1}{n_A},0,0\right),\nonumber\\
\vec f_2&=&\vec e_7=\left(\frac{1}{n_Ak},-\frac{1}{n_Ak},-\frac{1}{n_Bk},\frac{1}{n_Bk}\right),\nonumber\\
\vec f_3&=&\vec e_3-\vec e_1=(-1,0,1,0),\nonumber\\
\vec f_4&=&\vec e_1=(1,0,0,0).
\label{basisf}
\end{eqnarray}
Using this basis, the toric data become
\begin{equation}
\vec v_1=[0,0,0,1]_{\vec f},\quad
\vec v_2=[n_A,0,0,1]_{\vec f},\quad
\vec v_3=[0,0,1,1]_{\vec f},\quad
\vec v_4=[n_B,kn_B,1,1]_{\vec f},
\end{equation}
where $[a_1,\cdots,a_4]_{\vec f}=\sum_ia_i\vec f_i$.
We have chosen basis (\ref{basisf})
so that the toric data become the standard form
in which the last components of the vectors
are $1$.
We can draw the toric diagram using the first three components
of these vectors $\vec v_i$,
which coincides with that in Fig.~\ref{c4.eps}(b).

\section{Further generalization}\label{more.sec}
Up to now we have considered
a brane system with two kinds of fivebranes.
It is also possible to introduce more than two kinds of fivebranes.
To represent the types of branes we used $q_I=0$ and $1$.
In this section we allow $q_I$ to be an arbitrary integer.
In this case, we do not need to introduce the coefficient $k$
in (\ref{cscouplings}) and we set $k=1$.
This means that the $I$th fivebrane is a
$(q_I,1)$5-brane, and the Chern-Simons couplings are given by
\begin{equation}\label{kInonzero}
k_I=q_{I+1}-q_I.
\end{equation}
For simplicity we assume that the Chern-Simons couplings do not vanish.
This implies that all the adjoint chiral multiplets $\Phi_I$ and the
vector multiplets $V_I$ become massive.
It is easy to show that even if some of the $k_I$ vanish
we obtain the same moduli space as derived below.

By integrating out $\Phi_I$, we obtain the superpotential
\begin{equation}
W=-\sum_{I=1}^n\frac{1}{2(q_{I+1}-q_I)}(X_IY_I-Y_{I+1}X_{I+1})^2.
\end{equation}
(When we obtained the superpotential (\ref{ellw}) we used $q_I=0$ and $1$,
although we cannot use it here.)
From the assumption (\ref{coulomb}),
the F-term conditions for $X_I$ and $Y_I$ give
\begin{equation}
\frac{X_{I+1}Y_{I+1}-X_IY_I}{q_{I+1}-q_I}
=\frac{X_IY_I-X_{I-1}Y_{I-1}}{q_I-q_{I-1}},
\end{equation}
and this is solved as
\begin{equation}\label{XIYIaqIb}
X_IY_I=a+q_Ib,
\end{equation}
where $a$ and $b$ are arbitrary complex numbers.

The equations of motion of the auxiliary fields $\sigma_I$
are solved by (\ref{dtermc}) with the parameters $\lambda_I$ constrained
by (\ref{subg}).
The constraint (\ref{subg}) defines group $H$
and Eq.~(\ref{dtermc}) has the form of the D-term condition
associated with group $H$.
This group is identical to the continuous part of the
unbroken gauge symmetry,
which is given by (\ref{ubk}) with $k=1$.
The constraints imposed on the parameters are
\begin{equation}
\sum_{I=1}^n\lambda_I=0,\quad
\sum_{I=1}^nq_I\lambda_I\in\frac{1}{2\pi}{\mathbb Z}.
\label{hp}
\end{equation}

The following ``baryonic operators'' are invariant under
gauge symmetries satisfying (\ref{hp}),
\begin{equation}
x=\prod_{I=1}^nX_I,\quad
x_A=\prod_{I=1}^nX_I^{q_{\max}-q_I},\quad
x_B=\prod_{I=1}^nX_I^{q_I-q_{\min}},
\end{equation}
\begin{equation}
y=\prod_{I=1}^nY_I,\quad
y_A=\prod_{I=1}^nY_I^{q_{\max}-q_I},\quad
y_B=\prod_{I=1}^nY_I^{q_I-q_{\min}},
\end{equation}
where $q_{\min}$ and $q_{\max}$ are the minimum and maximum
of $q_I$, respectively.
Any gauge-invariant monomial of $X_I$ and $Y_I$ can be represented
as a monomial of the mesonic operators $M_I=X_IY_I$ and these baryonic operators.

We now have the following $8$ gauge-invariant variables:
\begin{equation}
a, b, x, x_A, x_B, y, y_A, y_B.
\end{equation}
These operators  satisfy
\begin{equation}\label{x0y0}
xy=\prod_{I=1}^n(a+q_Ib),\quad
x_Ay_A=\prod_{I=1}^n(a+q_Ib)^{q_{\max}-q_I},\quad
x_By_B=\prod_{I=1}^n(a+q_Ib)^{q_I-q_{\min}},
\end{equation}
\begin{equation}\label{x0x1x}
x_Ax_B=x^{q_{\max}-q_{\min}},\quad y_Ay_B=y^{q_{\max}-q_{\min}}.
\end{equation}
Because the first equation in (\ref{x0y0}) is
not independent of the other two
due to relations (\ref{x0x1x}),
these relations decrease the number of independent degrees of freedom by four,
and the moduli space becomes a complex 4-dimensional space.
If the $q_I$ take more than two different values,
the equations in (\ref{x0y0}) are not binary relations of monomials,
and the moduli space is nontoric.

\section{Discussion}
In this paper we studied the Higgs branch
of Maxwell-Chern-Simons theories
described by circular quiver diagrams.
We first considered the model realized by the D3-NS5-$(k,1)$5-brane
system with an arbitrary number of fivebranes,
and showed that the moduli space is the orbifold ${\mathbb C}^4/\Gamma$,
where $\Gamma$ is the discrete group
generated by (\ref{zna})--(\ref{znc}).
When we determined the orbifold group $\Gamma$,
we made an assumption for the flux quantization
(\ref{feqf2}).
Our result was confirmed by comparing it to
the M-theory dual of the brane configuration.
We also discussed the model realized by a brane system with
more than two kinds of fivebranes,
and we obtained a 4-dimensional nontoric moduli space.

Note that our result is different from that expected from
the orbifold method.
In general, a quiver gauge theory obtained by
the orbifold method introduced in Ref.~\citen{Douglas:1996sw}
includes $n$ copies of fields of the parent theory,
where $n$ is the order of the orbifolding group.
Such analysis is carried out in Ref.~\citen{Benna:2008zy} for the model
proposed in Ref.~\citen{Aharony:2008ug},
and a theory was obtained in which
the number of $U(N)$ factors in the gauge group
is proportional to the order of the corresponding orbifolding group.
On the other hand, our construction gives
$n=n_A+n_B$ copies of fields,
whereas the order of the orbifolding group is
proportional to the product $n_An_B$.
Because the brane construction and orbifold
method are both important methods for constructing field theories
in string theory,
it is very important to understand the reason for this
discrepancy.

The moduli spaces we obtained in this paper
are completely determined by the number
of fivebranes of each type.
In the case of the brane system with A- and B-branes,
the moduli space depends only on the level $k$ and
the numbers of fivebranes $n_A$ and $n_B$.
The orders of A- and B-branes along the compact direction
do not affect the moduli space.
This is also the case in the brane system with more than two
types of fivebranes discussed in \S\ref{more.sec}.
This may be interpreted as a duality similar to
the Seiberg duality in the 4-dimensional ${\cal N}_{(d=4)}=1$
supersymmetric gauge theories\cite{Seiberg:1994pq}.
In the 4-dimensional case,
this duality can be understood as the
exchange of the two types of branes\cite{Elitzur:1997fh}.
In the brane system we consider in this paper,
the exchange of A- and B-branes generates
new D3-branes by the Hanany-Witten effect\cite{Hanany:1996ie}.
It will be an interesting problem
to clarify the relation among
theories realized by brane systems
with different orders of fivebranes.

The models we considered in this paper
are expected to flow to conformal fixed points
in the low-energy limit,
and thus the AdS/CFT correspondence
is expected to be useful for studying low-energy dynamics.
When we discuss the AdS/CFT correspondence,
it is necessary
to establish the correspondence
between geometries and the UV description
of quiver gauge theories.
In the case of 4-dimensional ${\cal N}=1$ superconformal theories,
brane tiling\cite{Hanany:2005ve,Franco:2005rj,Franco:2005sm}
is a convenient tool for finding this correspondence in the toric Calabi-Yau case.
Although
the generalization of brane tiling
to 3-dimensional gauge theories has been proposed\cite{Lee:2006hw,Lee:2007kv,Kim:2007ic},
much less is known about the duality
in the 4-dimensional case
due to the small number of examples.
We hope that the examples in this paper will be useful for
investigating the general relation between four-manifolds
and quiver Chern-Simons gauge theories.

\section*{Acknowledgements}
We would like to thank T.~Eguchi for valuable discussions.
We would also like to acknowledge the helpful comments of K.~Ohta.
Y.~I. is partially supported by
a Grant-in-Aid for Young Scientists (B) (\#19740122) from the Japan
Ministry of Education, Culture, Sports,
Science and Technology.

%

\end{document}